# Magnetic Field-Tunable Repulsive Exciton–Exciton Interaction in the van der Waals Antiferromagnet $NiPS_3$

Kaiyang Huang [1], Jaena Park[2]; Zhuo Yang[1]; Je-Geun Park[2]; Atsuhiko Miyata [1], Yoshimitsu Kohama[1], Yasuhiro H. Matsuda[1*]

[1]The Institute for Solid State Physics, The University of Tokyo, Kashiwa, Chiba 277-8581, Japan

[2]Department of Physics and Astronomy, Seoul National University (SNU), Seoul 08826, Republic of Korea

**ABSTRACT**. Two ultra-narrow absorption peaks around 1.5 eV, which are widely believed to originate from a transition from a spin-orbital entangled triplet to a singlet state, in the two-dimensional van der Waals crystal $NiPS_3$, have attracted considerable attention because of their pronounced spin-dependent character. An interesting question is whether ultrahigh magnetic fields modify an interaction-driven hybridization between those two peaks. In this work we perform systematic magneto-optical measurements of $NiPS_3$ in pulsed magnetic fields of up to 178 T and observe a pronounced mutual repulsion between the two sharp exciton peaks accompanied by a redistribution of oscillator strength, while the band edge shows no detectable field-induced shift within our experimental resolution. We construct a minimal two-level interaction model and compare it semi-quantitatively with the experimental data. Our results reveal a magnetic-field-tunable exciton–exciton coupling as the dominant high-field response of $NiPS_3$, and clarify this material as a new experimental platform for exploring strongly correlated exciton physics in magnetic van der Waals insulators.

## Introduction

Two-dimensional (2D) magnetic van der Waals materials provide a highly promising platform for addressing fundamental questions in condensed-matter physics [1][2]. Among them, $NiPS_3$ has emerged as a prototypical system [3][4]. $NiPS_3$ is a layered honeycomb-lattice $S = 1$ antiferromagnet with the Néel temperature $T_N \approx 155$ K; it exhibits zigzag magnetic order and easy-plane (XY-type) anisotropy [3][4], making it an excellent test bed for investigating the origin of 2D magnetism and magnetic anisotropy. In this material, two ultra-narrow, strongly spin-dependent near-infrared excitonic resonances at ~1.475 eV and ~1.50 eV have attracted particular interest [5][6][7][8]. Most previous studies attribute the lower-energy peak at 1.475 eV to a triplet-to-singlet excitation, often described as a Zhang–Rice-like transition [5][6][9], arising from Ni 3*d*–S 3*p* hybridization and analogous to the Zhang–Rice states first identified in cuprates [10]. Other recent theoretical works have instead interpreted the same low-energy excitation within a Hund's-coupling-based picture [11][12]. However, both viewpoints are in broad agreement regarding its triplet-to-singlet character, whereas the microscopic origin of the higher-energy peak remains unsettled. [5][6][9][11][12]. The magneto-optical experiments have revealed a clear Zeeman splitting of the low-energy exciton when the magnetic field is applied along the easy *a* axis, together with a spin-flop transition around 10 T [9][13]. Just above the spin flop, the spins tend to align nearly perpendicular to the external field, and the field variation of the exciton peak position at higher fields becomes similar to that when a field is applied along the $c^*$ axis [9][13]. For fields along $c^*$, only a small energy shift is observed [6][13]. This behavior has been interpreted in terms of interactions between bright and dark excitons [13][14][15][16][17][18]. However, the absence of systematic measurements in higher magnetic fields has so far prevented a clear separation of Zeeman, diamagnetic, and exciton–exciton interaction contributions.

In this work, we perform a detailed magneto-optical absorption study of $NiPS_3$ in ultrahigh magnetic fields. At temperatures down to 1.4 K and in fields up to 178 T, we resolve a pronounced mutual repulsion between the two sharp excitonic resonances accompanied by a redistribution of oscillator strength, while the band edge shows no detectable shift within our experimental resolution. Such magnetically tunable excitons are evidence of a close link between the two excitonic branches. $NiPS_3$ thus provides a field-tunable platform for exciton–exciton interactions and will facilitate further fundamental studies of two-dimensional magnets.

The 2D van der Waals insulator $NiPS_3$ crystallizes in a monoclinic structure with space group C2/m [3][4][19]. At low temperatures it exhibits zigzag antiferromagnetic order within the *ab* plane (Fig. 1(a))[3][4][19][20][21]. In this monoclinic setting, the layers lie in the *ab* plane, and the reciprocal-lattice axis $c^*$ is strictly perpendicular to the *ab* plane, whereas the real-space *c* axis is slightly tilted away from this normal. In our plate-like crystals the surface normal is therefore parallel to $c^*$. This quasi-two-dimensional antiferromagnetic honeycomb lattice is naturally described by an XXZ-type spin model [3]. As shown in Fig. 1(a), the Ni spins form an $S = 1$ sublattice with antiparallel alignment along the *a* axis, while in the *ac* plane, the Ni spin direction is not strictly aligned with the *a* axis, instead, it forms an angle of about 8° with the *a* axis [4][19][21]. This ordered state is destroyed upon heating through the $T_N$ ≈ 155 K [3][4][19].

*Contact author: ymatsuda@issp.u-tokyo.ac.jp

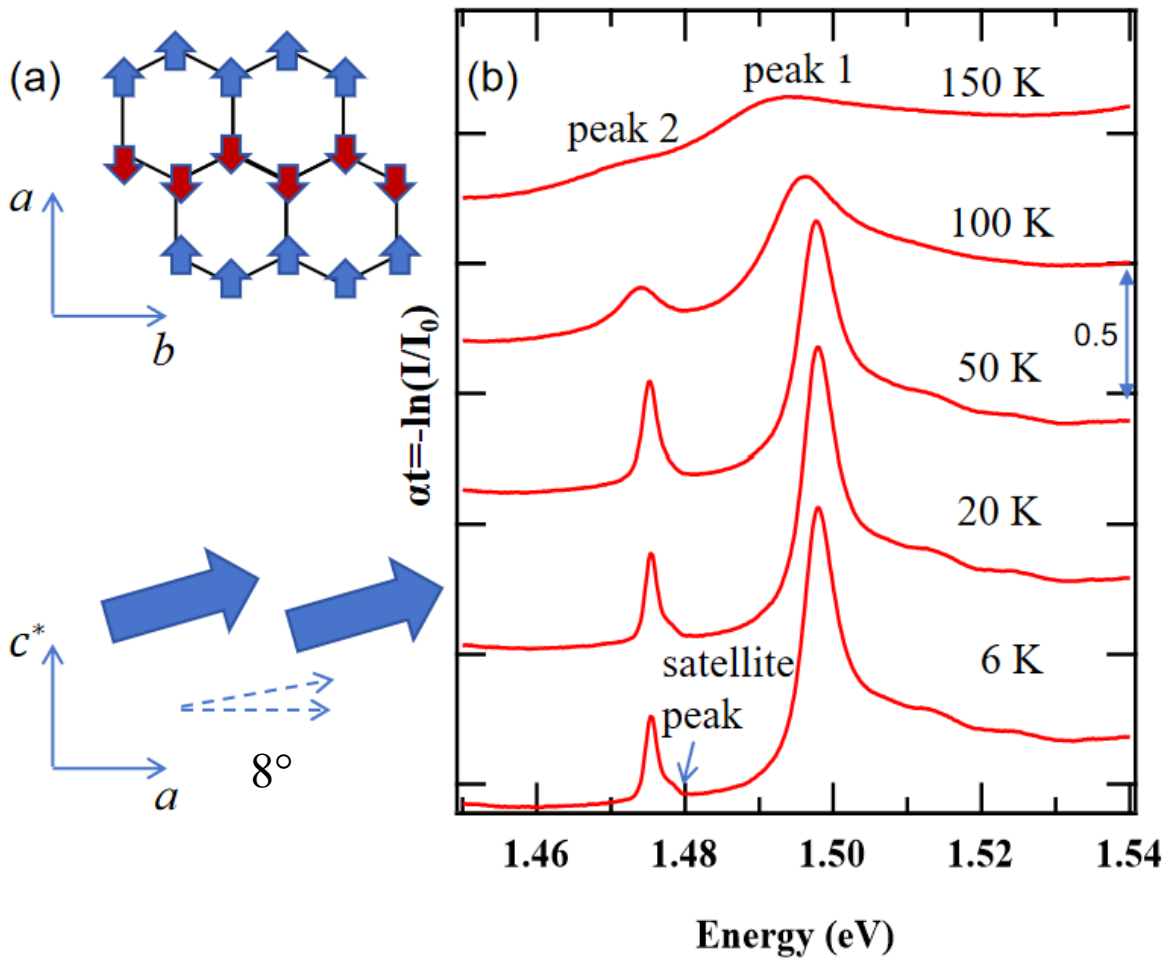

**FIG. 1**(a) Magnetic structure of $NiPS_3$. In the *ab* plane the $Ni^{2+}$ moments form a zigzag antiferromagnetic pattern with spins aligned along the parallel and anti-parallel directions of the crystallographic *a* axis. In the *ac* plane the spins of Ni are tilted by about 8° away from *a* toward *c**. (b) Zero-field absorption spectra of $NiPS_3$ at selected temperatures. Spectra are vertically offset by 0.5 for clarity. Below 50 K the two sharp excitonic peaks (peak 1 and peak 2) show only minor changes, whereas on approaching the $T_N$ ≈ 155 K, they considerably weaken and their linewidths broaden, indicating a strong link between these excitons and the antiferromagnetic order. As for peak 2, a satellite peak is observed at low temperatures.

## Experiment

For our optical measurements, we exfoliate layered $NiPS_3$ samples using adhesive tape and mount them on 2-mm-diameter quartz plates using cryogenic glue; the typical sample thickness is a few tens of μm. Ultrahigh magnetic fields of up to 178 T are generated by a destructive single-turn coil [22][23]. A Xe-flash lamp is employed as the broadband light source and a streak camera is used to record the optical absorption within a ~10 μs time window during the field pulse [24][25]. Two optical fibers (800 μm core diameter) are employed for irradiation and detection of light. The plate-wise sample is sandwiched between two optical fibers for the optical transmission measurements. The light direction is parallel to a magnetic field orientation. In addition, we perform complementary measurements in a non-destructive pulsed magnet [26] providing fields up to 52 T; under these conditions, we use a longer effective exposure (integration) time to obtain the absorption spectra with a higher signal-to-noise ratio[27].

## Results of the experiment

The zero-field absorption spectrum at 6 K (Fig. 1(b)) shows a sharp peak at 1.498 eV (peak 1) and a second sharp peak at 1.475 eV (peak 2), both with linewidths of nearly 10 meV. The optical absorption is expressed by αt, where α and t represent the optical absorption coefficient of $NiPS_3$ and the sample thickness, respectively. αt is experimentally deduced by $-\ln(I/I_0)$ where I and $I_0$ are the transmitted and original (reference) light intensity, respectively. As the temperature is increased, both absorption peaks gradually shift to lower energies and broaden, and eventually disappear when the temperature approaches the $T_N$. These behaviors are consistent with previous reports [5][6][7].

To further investigate how these two sharp absorption features respond to ultrahigh magnetic fields, we performed systematic pulsed-field magneto-absorption measurements on $NiPS_3$. At 10 K, magnetic fields of up to 178 T were applied along the crystallographic $c^*$ axis of $NiPS_3$. The exciton energy peak shifts are converted to the shifts in the photon energy and plotted as a function of magnetic field in Fig. 2. The higher-energy absorption peak around 1.498 eV (peak 1) shows a clear shift toward higher energy with increasing magnetic fields. In contrast, the lower-energy exciton peak around 1.475 eV (peak 2) shifts to lower energy with increasing magnetic fields. It is found that the peak shifts are clearly nonlinear in low fields, while at higher fields the field variation of both branches becomes approximately linear in $B$. The original absorption spectra are shown in End Matter.

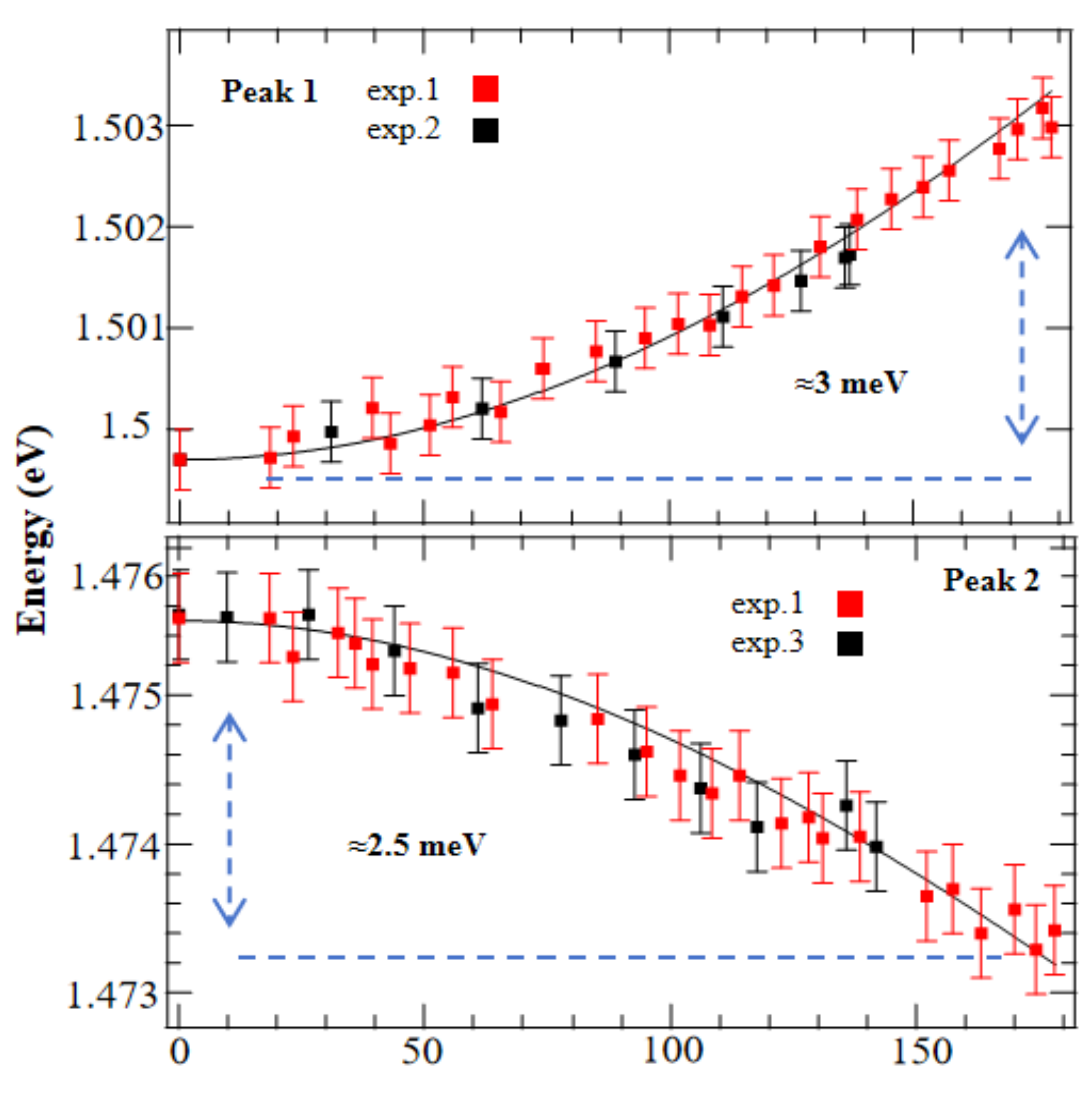

**FIG. 2** Field dependence of the exciton peak energies. Red symbols denote the data of the experiment up to 178 T(exp.1). The uncertainty in the peak position is ±0.3 meV for each point. Black symbols denote the data of another experiments (exp. 2 and exp.3) up to 140 T. The black curves are fits to the two-level interaction model. At 10 K and 178 T, the higher-energy exciton shifts upward by about 3 meV, while the lower-energy exciton shifts downward by about 2.5 meV.

*Contact author: ymatsuda@issp.u-tokyo.ac.jp

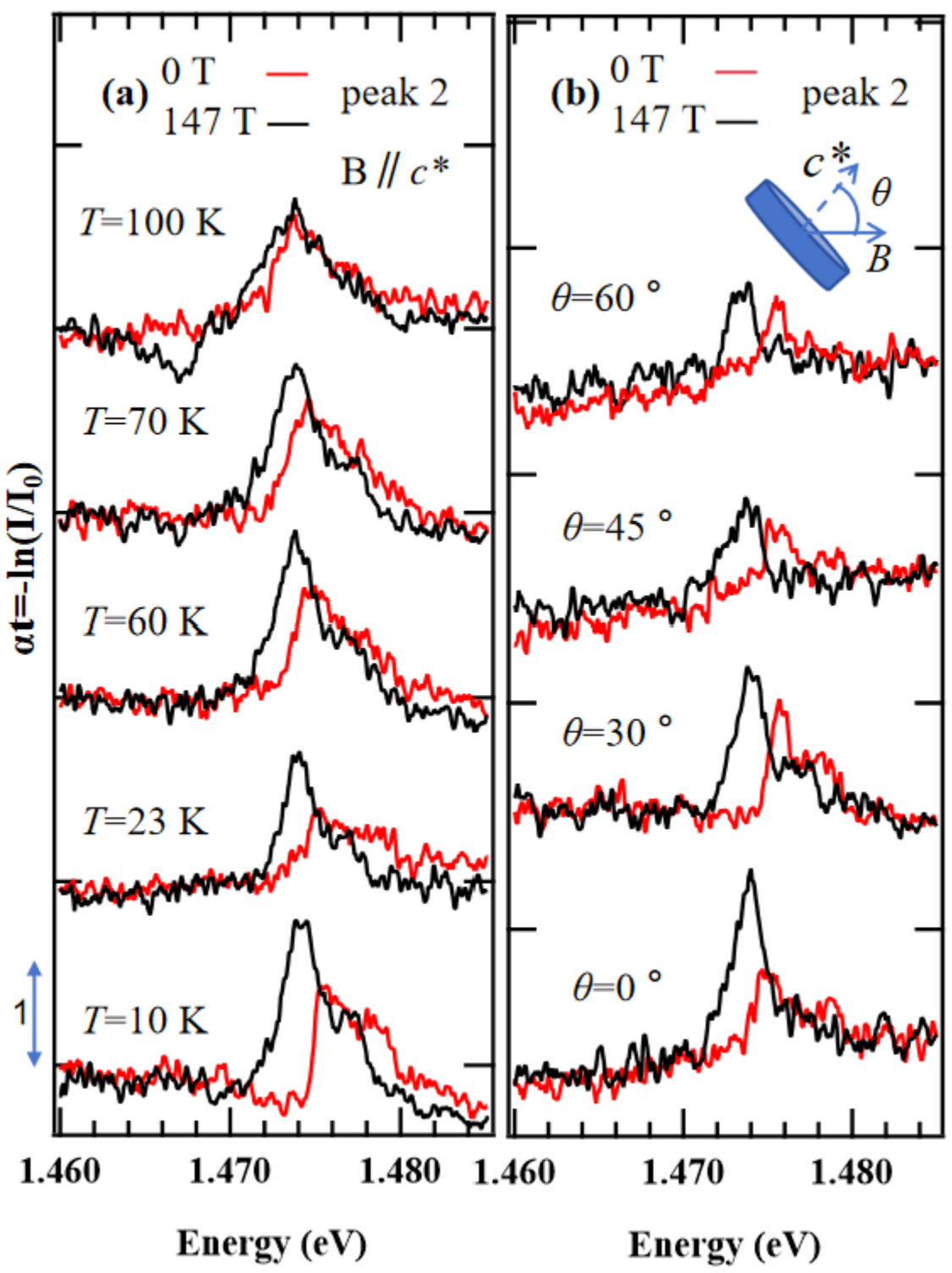

**FIG. 3**(a) Temperature dependence of the spectra near peak 2 for zero magnetic fields (red curve), and magnetic fields of 147 T applied along the $c^*$ axis (black curve). Spectra are vertically offset by 2 for clarity. (b) The field orientation dependence of the spectra near peak 2 at 10 K are shown for the sample is tilted so that a magnetic field angle from the $c^*$ axis ($\theta$) equal to 30°, 45°, and 60°, respectively. The slight change in the zero field spectra (red curves) for different $\theta$ is mainly due to the change in the optical path by tilting the sample. Spectra are vertically offset by 2 for clarity.

To clarify the origin of the distinct peak energy shifts at high fields, we analyzed the spectra of similar experiments at different temperatures up to 147 T. Fig. 3(a) shows the spectra of peak 2 at zero and 147 T for each temperature. The weak side peak on the low-energy side of peak 2 is assigned to the phonon-related satellite reported in Ref. 6; it follows the field shift of peak 2 and is not treated as an independent field-induced feature. The peak position is shifted toward low energy by the application of 147 T at each temperature. A summary of the extracted peak shift versus temperature is given in the End Matter (Appendix, Fig. 6(a)). We find that the magnitude of the shift is nearly temperature independent well below the $T_N$ but is strongly reduced as the temperature approaches $T_N$. The spectra are analyzed with a curve fitting, and the peak positions are evaluated. (The details of the curve fitting analysis are shown in the Supplementary material[28].) Note that we have used circularly polarized light with opposite helicities in another experiment and found no systematic dependence of the field-induced shift on the polarization state of the incident light within our experimental accuracy.

To examine the field-orientation dependence in the high-field regime, we measured spectra at 147 T with the sample tilted away from the $c^*$ axis by $\theta$ = 30°, 45°, and 60°. The tilt-angle dependence of the field-induced shift is shown in the End Matter [Appendix, Fig. 6(b)]. Within our experimental resolution, the shifts show little angular dependence [Fig. 3(b)]. This does not contradict Ref. 9, where a distinct field angle dependence of the peak position is observed at the field below the spin-flop transition field. After the spin-flop transition, the spins become nearly perpendicular to the field and then gradually cant toward it with increasing field, making the response qualitatively similar to that for $B$ // $c^*$. In addition, because the earlier measurements were limited to lower fields and did not follow the full evolution of both branches, especially the higher-energy one, the opposite strong-field-induced shifts were not identified there.

## Discussion

Previous work has reported that peak 2 exhibits a slight shift towards lower energy by a magnetic field and has attributed this behavior to coupling between bright and dark excitons[9][13]. In our measurements, however, the satellite feature on the low-energy side follows essentially the same field dependence as the main peak as shown in Fig. S5 in the Supplementary material, and we do not observe any substantial exchange of intensity between them [28].

Other studies [29][30] have instead invoked a diamagnetic shift to account for the high-field behavior, but in general a diamagnetic contribution tends to push exciton levels to higher energy. Taken together, these observations indicate that bright–dark exciton mixing or diamagnetic effects alone cannot account for the magnetic field-induced red shift of peak 2 in the present work.

We interpret the opposite field shifts of the two excitons in Fig. 2(b) as level repulsion between two closely related excitonic states[31][32][33]. Ref. 11 provides a possible microscopic basis: the two peaks can be viewed within a common two-orbital manifold as nearly degenerate states split by lowered symmetry. In this picture, the magnetic field can primarily enhance the mixing of the two electronic states through the change in the antiferromagnetic states rather than rigidly shifting the band structure. This is consistent with the opposite motion of the two narrow peaks, the nearly unchanged band edge, and the weak redistribution of oscillator strength. This behavior can be captured by a simple two-level model. In our minimal two-level model, the coupled excitons are described by the Hamiltonian.

*Contact author: ymatsuda@issp.u-tokyo.ac.jp

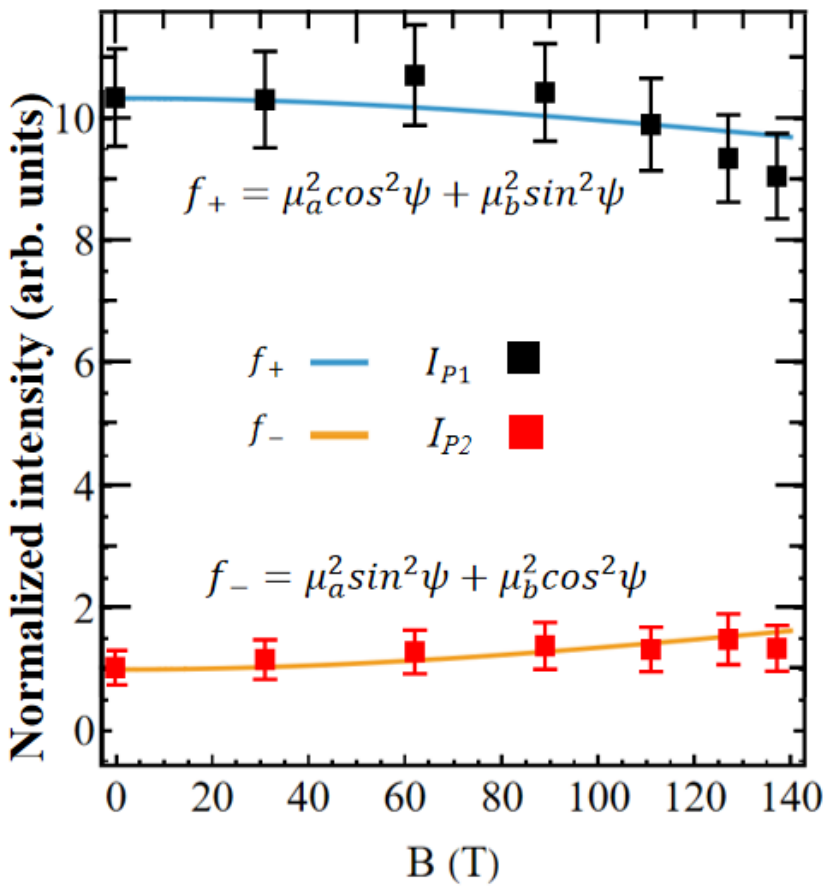

**FIG. 4** The magnetic field dependence of the normalized exciton absorption intensity of peak 1 ($I_{P1}$) and peak 2 ($I_{P2}$) are shown with closed squares. The exciton intensities are approximated by the peak area, and are normalized to the zero-field intensity of the low-energy peak $I_{P2}$ (0 T), such that $I_{P2}$ (0 T) = 1. The calculated oscillator-strength transfer between the two excitonic resonances within the framework of a two-level model, $f_+$ ($B$) and $f_-$ ($B$), are shown by solid curves. Their initial values are set as $f_+(0)=I_{p1}(0)$, $f_-(0)=I_{p2}(0)$.

$$\mathrm{H}=\begin{bmatrix} E_1+\beta B^2 & V(B) \\ V(B) & E_2+\beta B^2 \end{bmatrix}, \quad (1)$$

where $E_1$ and $E_2$ are the zero-field energies of the two excitons, and the diagonal term $\beta B^2$ represents a phenomenological diamagnetic contribution that produces a common upward shift of both levels. The off-diagonal term $V(B)$ represents field-induced hybridization between the two excitons on the antiferromagnetic background. Motivated by an off-diagonal Zeeman-like coupling, we assume that it is linear in $B$ and write

$$V(B)=\frac{1}{2}g_{in}\mu_B B, \quad (2)$$

where $g_{in}$ is an effective coupling $g$ factor and $\mu_B$ is the Bohr magneton. In this effective description, $V(B)$ phenomenologically describes magnetic-field-induced mixing between two nearby excitonic states, consistent with an off-diagonal Zeeman-like coupling. Diagonalizing this Hamiltonian yields the field-dependent eigenvalue

$$E_\pm=\frac{1}{2}(E_1+E_2)+\beta B^2\pm\sqrt{(\frac{E_1-E_2}{2})^2+(\frac{g_{in}\mu_B B}{2})^2}\,. \quad (3)$$

From the expression for $E_\pm(B)$, the field-induced energy shifts are expected to be purely quadratic in $B$ at low fields, whereas at intermediate and higher fields (still well below magnetization saturation) the linear-in-$B$ term dominates. Using this formula, we fit the experimental energy shifts of the two excitonic peaks at 10 K over the full field range up to 178 T (Fig. 2) with a single set of parameters and find that it well reproduces both the low-field curvature and the approximately linear behavior at higher fields. The parameters used for the fitting are summarized in Table S1 in the Supplementary material [28]. In addition, as the magnetic field increases, we observe a weak but systematic transfer of oscillator strength from peak 1 (higher-energy peak) to peak 2 (lower-energy peak) as shown in Fig. 4. Representative spectra at 0 T and 137 T illustrating this redistribution are provided in the End Matter (Appendix, Fig. 7). The relative peak intensity between peak 1 and peak 2 is found to be $I_{p1}$:$I_{p2}$ = 10.3:1 at zero fields. It is found that $I_{p1}$ and $I_{p2}$ change with increasing field, indicating the transfer of the oscillator strength. This small redistribution is consistent with a finite interaction between the two excitons and provides further support for our interpretation in terms of a coupled two-level system, in line with standard two-level mixing and oscillator-strength redistribution scenarios discussed in exciton theory [34][35][36][37]. The eigenstate of this two-level model can be written as

$$\begin{aligned} |+\rangle &= \cos(\psi)|a\rangle+\sin(\psi)|b\rangle \\ |-\rangle &= \sin(\psi)|a\rangle+\cos(\psi)|b\rangle \end{aligned}\,. \quad (4)$$

Here, $|a\rangle$ and $|b\rangle$ denote two closely related excitonic basis states from a common microscopic manifold. And $\psi$ is a mixing angle that parametrizes the field-induced hybridization of the two levels. Within our two-level Hamiltonian, $\psi$ is determined by

$$\tan(2\psi)=\frac{g_{in}\mu_B B}{E_1-E_2}\,. \quad (5)$$

*Contact author: ymatsuda@issp.u-tokyo.ac.jp

Within this framework, the evolution of the oscillator strengths $f_\pm$ of the two eigenmodes can be written as follows.

$$\begin{aligned} f_+ &= \langle + | + \rangle \\ f_- &= \langle - | - \rangle \end{aligned} \tag{6}$$

In this situation, as the external magnetic field is increased, the oscillator strengths of the two excitons mutually compensate each other, such that a corresponding enhancement of the other offsets the reduction in the intensity of one branch. At the same time, the mixing angle $\psi$ remains relatively small, reaching only about 18.5°even at 178 T. As a result, the exchange of oscillator strength between the two branches is correspondingly weak, and as shown in Fig. 4, it is in agreement with the small intensity changes observed experimentally. The larger deviation for peak 1 likely reflects the greater uncertainty in evaluating its modest field-induced intensity change on a structured background, as well as the limitation of the minimal two-level model, which does not include weak satellite features as additional spectral-weight channels.

To understand how this interaction evolves with temperature, we proceed with the following picture. Previous studies have established that both exciton peaks are closely tied to the antiferromagnetic state of $NiPS_3$ [5][6][7][8]; as the temperature approaches the $T_N$, the peaks broaden, lose intensity, and eventually disappear once the temperature exceeds $T_N$. We therefore regard the interaction energy between the excitons as being controlled by the antiferromagnetic order parameter, i.e., by the Néel vector, similar in spirit to other antiferromagnetic semiconductors where magnetic dynamics modulate exciton–exciton interactions [32]. The temperature dependence of the magnetization at a low magnetic field around 1 T in $NiPS_3$ is shown in Fig. S6 in the Supplementary material [28]. Over a wide temperature range well below $T_N$, the magnetization changes very little, indicating that the antiferromagnetic order is essentially intact. At a fixed strong external field exceeding around 147 T, however, when the temperature is increased from far below $T_N$ toward $T_N$, the magnetization is expected to rise sharply, signaling that long-range antiferromagnetic order is strongly suppressed by thermal fluctuations. As a result, the magnetic susceptibility is enhanced; under the same field, the induced magnetization increases substantially, and the system's response gradually crosses over from that of a "rigid antiferromagnet" to a more paramagnetic-like state-where the Néel vector is expected to be diminished. Our experimental results follow the same trend [Fig. 3(a)]: Far from $T_N$, the field-induced energy shifts of the two excitons are essentially independent of temperature, whereas they are strongly reduced when the temperature approaches the $T_N$.

Although a strong magnetic field can in principle make the spins cant away from the easy axis and thereby reduce the projection of the Néel vector, the exciton energy shifts analyzed here are obtained for fields applied along the hard $c^*$ axis (or some tilted angles from the $c^*$ axis), where the dominant effect is to induce a finite uniform magnetization rather than to completely quench the antiferromagnetic order. According to the theoretical paper [38], even at 178 T, the system is calculated to be far from magnetic saturation. Consequently, the spin–orbital-entangled excitonic configuration that governs the off-diagonal Zeeman coupling between the two excitons remains essentially unchanged, as reflected by the fact that the line shapes of the two peaks stay intact even at 178 T. This robustness is consistent with the effective coupling $g_{in}$ factor extracted from the 0 – 178 T range remains constant value (around 2) and thus supports the above picture.

## Conclusions

In conclusion, we have experimentally demonstrated an effective repulsive interaction between the two narrow-linewidth excitons in the van der Waals insulator $NiPS_3$. In ultrahigh magnetic fields, the two excitonic branches move apart in energy while the band edge remains essentially unchanged, and a finite yet modest redistribution of oscillator strength occurs between them. Together with the two-level interaction model, these results establish $NiPS_3$ as a field-and temperature-tunable platform for exciton–exciton interactions on an antiferromagnetic background, providing a basis for understanding and controlling strongly correlated excitonic states and for exploring their potential device applications [39][40]. At the same time, the robustness of the $NiPS_3$ excitons against very high magnetic fields highlights their promise as probes or sensors in electronic as well as optical under extreme conditions.

## Acknowledgements

This work was funded by the JSPS KAKENHI, Grant-in-Aid for Transformative Research Are-as (A) Nos.23H04859 and 23H04860, Grant-in-Aid for Scientific Research (B) No.23K25814. The work at SNU was funded by the Leading Researcher Program of the National Research Foundation of Korea (Grant No. RS-2020-NR049405). Z.Y. was supported by research grant from Research Foundation for Opto-

*Contact author: ymatsuda@issp.u-tokyo.ac.jp

Science and Technology and JSPS KAKENHI Grant Number 25K17324 (Grant-in-Aid for Early-Career Scientists).

Data Availability - The data that support the findings of this Letter are available [41].

*Contact author: ymatsuda@issp.u-tokyo.ac.jp

*Contact author: ymatsuda@issp.u-tokyo.ac.jp

# End Matter

*Appendix: Additional spectra and robustness checks.* Figures 5–7 provide supporting data and consistency checks for the main-text analysis.

The two-dimensional absorption spectra are shown in Fig. 5. The vertical cross-section gives the absorption spectrum at each magnetic field ($B$), and the spectra at different $B$ are shown in Fig. S4 in the supplementary material. The higher-energy absorption peak around 1.498 eV (peak 1) before approximately 4 μs corresponds to the zero-field peak position and shows a clear shift toward shorter wavelength at approximately 6.2 μs, where a magnetic field reaches 178 T (blue shift). As for the lower-energy exciton peak around 1.475 eV (peak 2), it shifts to longer wavelength with increasing magnetic fields (red shift).

The temperature-dependent field-induced red shift of peak 2 at 147 T is shown in Fig. 6 (a). And, the field angle ($\theta$) variation of the evaluated peak shifts at 147 T is shown in Fig. 6 (b)

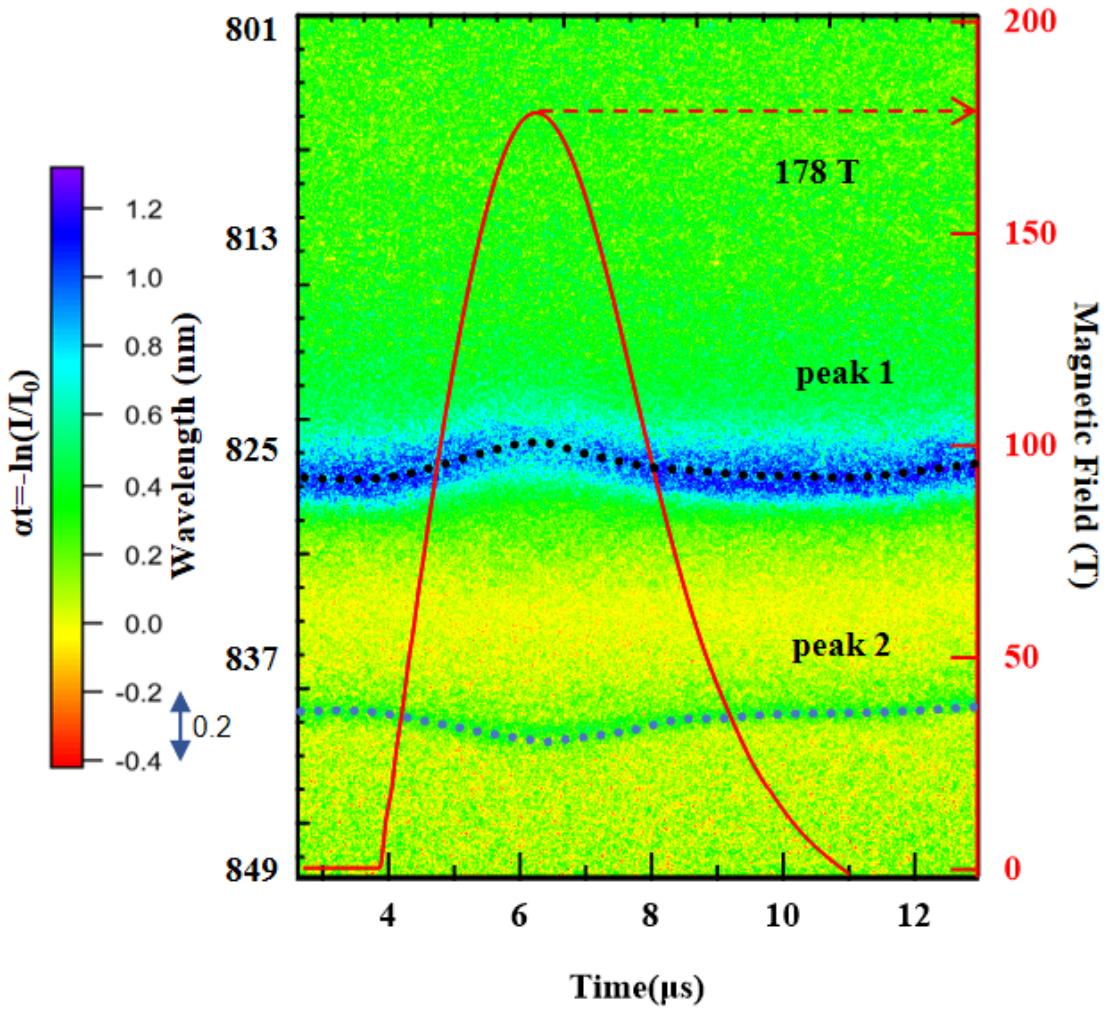

**FIG. 5** Magneto-absorption spectra of $NiPS_3$ in the vicinity of the peak 1 and peak 2 for magnetic fields up to 178 T applied along the $c$* axis at 10 K.

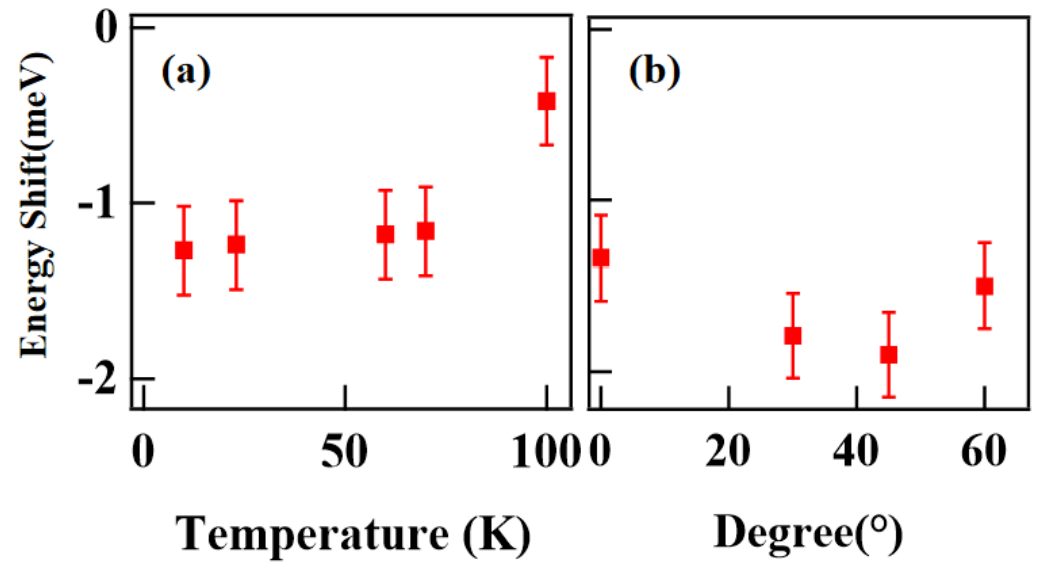

**FIG. 6** (a)A plot of the temperature dependence of the field-induced energy shift of peak 2 at 147 T. (b) The field angle dependence of the field-induced peak energy shift at 10 K and 147 T.

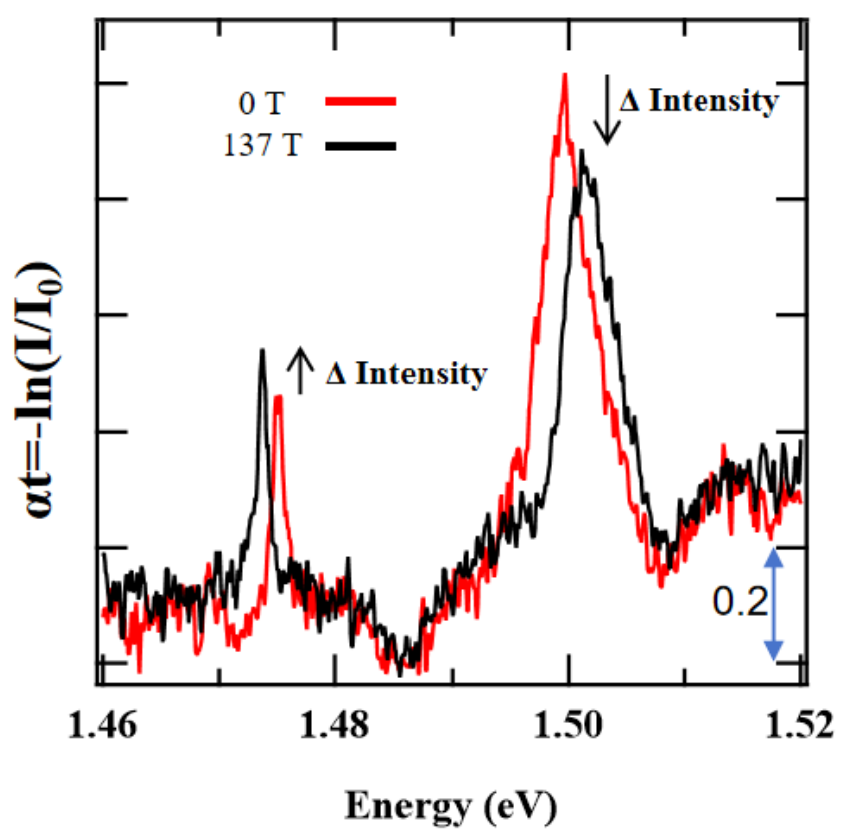

**FIG. 7** The absorption spectra at zero field and 137 T at 10 K.

*Contact author: ymatsuda@issp.u-tokyo.ac.jp